\documentclass[aps,pre,twocolumn,groupedaddress,showpacs]{revtex4}%
\usepackage{graphicx}
\usepackage{amsmath}
\usepackage{amsfonts}
\usepackage{amssymb}
\begin{document}

\title{Noise Enhanced Stability in Fluctuating Metastable States}
\author{Alexander A. Dubkov$^{1}$, Nikolay V. Agudov$^{1}$, and Bernardo Spagnolo$^{2}$}
\affiliation{$^{1}$Radiophysics Department, Nizhni Novgorod State
University, 23 Gagarin ave., 603950 Nizhni Novgorod, Russia}
\affiliation{$^{2}$INFM and Dipartimento di Fisica e Tecnologie
Relative, Group of Interdisciplinary Physics, Universit\`{a} di
Palermo, Viale delle Scienze, I-90128 Palermo, Italia}
\date{\today}

\begin{abstract}
We derive general equations for the nonlinear relaxation time of
Brownian diffusion in randomly switching potential with a sink.
For piece-wise linear dichotomously fluctuating potential with
metastable state, we obtain the exact average lifetime as a
function of the potential parameters and the noise intensity. Our
result is valid for arbitrary white noise intensity and for
arbitrary fluctuation rate of the potential. We find noise
enhanced stability phenomenon in the system investigated: the
average lifetime of the metastable state is greater than the time
obtained in the absence of additive white noise. We obtain the
parameter region of the fluctuating potential where the effect can
be observed. The system investigated also exhibits a maximum of
the lifetime as a function of the fluctuation rate of the
potential.

\end{abstract}
\maketitle

\section{Introduction}

Activated-escape in systems with metastable states underlies many
physical, chemical and biological problems. Examples are crystal
growth, tunnel diode, lasers, quantum liquids, spin systems,
protein folding and polymer physics \cite{Leg84,Man00}. The most
interesting and stubborn case of metastable state is one described
by time dependent potential, which fluctuates on a characteristic
time scale that may vary over a large range. In particular, the
metastable states with fluctuating barriers are common to chemical
and biological models \cite{Mie00,Bie93,Gri83}, and to a wide
range of physical problems, such as nonequilibrium transport
models, molecular dissociation in strongly coupled chemical
systems \cite{Mad92}, ratchet models for the action of molecular
motors \cite{Rei02}, noise in microstructures and generation
process of carrier traps in semiconductors
\cite{Sme99,Agu01,Doe94}. The escape from metastable state with
fluctuating or randomly switching barrier was studied in the past
mainly by well known mean first passage time (MFPT) technique. In
Refs.~\cite{Bal88,Doe92} exact results for the MFPT of escape
process over fluctuating barrier that switches between two
configurations have been obtained. However, the MFPT method
requires the implication of absorbing boundary in the system, and
it does not take into account the inverse probability current
through this boundary. The nonlinear relaxation time (NLRT) method
is devoid of this disadvantage \cite{Agu99}. Nevertheless the
theory for the NLRT is not well developed and the equations for
the NLRT are unknown for the case of time varying potential.\\
\indent In the present paper we derive general equations for the
NLRT for potentials randomly switching between two arbitrary
configurations with a sink. We find the exact solution of these
equations for a piece-wise linear potential flipping between
unstable and metastable configurations, for arbitrary white noise
intensity and fluctuation rate of the potential. Analyzing this
exact result we focus on the noise enhanced stability (NES)
effect, which implies that the system remains in the metastable
state for a longer time than in the absence of additive white
noise, and the lifetime of the metastable state has a maximum at
some noise intensity. This effect, which cannot be described by
Kramers-like behavior, was observed and investigated theoretically
and experimentally in various physical systems and mainly
concerning MFPT in periodically or randomly driven metastable
states
\cite{Mie00,Agu01,Man96,Agu98,Hir82,Day92,Mal96,Apo97,Wac99,Dan99,Yos03,Xie03,Spa04}.
In these papers the nonmonotonic behavior of the average escape
time was observed: (i) in physical systems, like tunnel diode
\cite{Man96}, and Josephson junction \cite{Mal96}, where the
influence of thermal fluctuations on the superconductive state
lifetime  and the turn-on delay time for a single Josephson
element with high damping was investigated; (ii) in chemical
systems, like the one-dimensional return map of the
Belousov-Zhabotinsky reaction, by investigating the behavior of
the lenght of the laminar region as a function of the noise
intensity \cite{Yos03}, and (iii) in biologically motivated
models, such that investigated in ref \cite{Mie00}, where the
overdamped motion of a Brownian particle moving in an asymmetric
fluctuating potential shows noise induced
stability.\\
\indent Here we study the NES phenomenon for the NLRT in randomly
switching metastable state, and we obtain analytically the region
of system parameters, where this effect takes place. We find also
resonant activation phenomenon by investigating the mean lifetime
as a function of switchings mean rate. Moreover, we find that the
NLRT exhibits a maximum as a function of barrier switching rate.
This new resonant-like phenomenon is related to the NES effect
\cite{Agu01,Man96}.\\
\indent The paper is organised as follows. In the second section
we derive the general equations for the nonlinear relaxation time
of Brownian diffusion in randomly switching potential with a
metastable state. In the third section we analytically derive the
mean lifetime for piece-wise linear potential. In the next section
we obtain the condition to observe the NES phenomenon and
investigate the behavior of the mean life time as a function of
switchings mean rate. In the final section we draw the
conclusions.
\section{General equations}

\begin{figure}[ptb]
\includegraphics{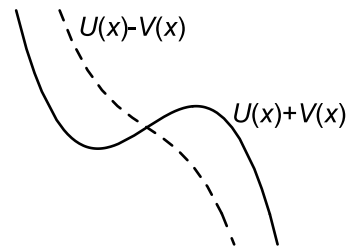}\caption{Switching potential with metastable state.}%
\label{fig1}%
\end{figure}

We consider the one-dimensional overdamped Brownian motion in switching
potential profile
\begin{align}
\frac{dx}{dt}  &  =-\frac{\partial\Phi\left(  x,t\right)  }{\partial x}%
+\xi\left(  t\right) ,\nonumber\\
\Phi\left(  x,t\right)   &  =U\left(  x\right)  +V\left(  x\right)
\eta\left(  t\right)  . \label{Lang}%
\end{align}
Here $x\left(  t\right)  $ is the Brownian particle displacement, $\xi(t)$ is
the white Gaussian noise with zero mean and correlation function $\langle
\xi\left(  t\right)  \xi\left(  t+\tau\right)  \rangle=2D\delta\left(
\tau\right)  $. The potential $\Phi\left(  x,t\right)  $ is the sum of two
terms: the fixed potential $U\left(  x\right)  $ and the randomly switching
term $V\left(  x\right)  \eta\left(  t\right)  $. The variable $\eta\left(
t\right)  $ is the Markovian dichotomous noise, which takes the values $\pm1$
with the mean flipping rate $\nu$. If we invoke the following expression for
probability density in terms of the average
\begin{equation}
W\left(  x,t\right)  =\langle\delta\left(  x-x(t)\right)  \rangle\label{known}%
\end{equation}
and introduce auxiliary function $Q\left(  x,t\right)  $
\begin{equation}
Q\left(  x,t\right)  =\langle\eta\left(  t\right)  \delta\left(
x-x(t)\right)  \rangle, \label{qu}%
\end{equation}
we obtain the next closed set of equations (see \cite{Luc92,Dub03}
and Appendix)
\begin{align}
\frac{\partial W}{\partial t} =\frac{\partial}{\partial x}\left[
U^{\prime}\left(  x\right)  W+V^{\prime}\left(  x\right) Q\right]
+D\frac{\partial^{2}W}{\partial x^{2}},\nonumber\\
\frac{\partial Q}{\partial t} =-2\nu Q+\frac{\partial}{\partial
x}\left[ U^{\prime}\left(  x\right)  Q+V^{\prime}\left(  x\right)
W\right]
+D\frac{\partial^{2}Q}{\partial x^{2}}. \label{two}%
\end{align}
Let $x=x_{0}$ be the initial position of Brownian particles. Then
\begin{equation}
W\left(  x,0\right)  =\delta\left(  x-x_{0}\right)  , \label{ini}%
\end{equation}
and $W\left(  x,t\right)  $ becomes the conditional probability
density $W\left(  x,t\left\vert x_{0},0\right.  \right)  $. Since
$\eta\left( 0\right)  $ is a deterministic value, the initial
condition for the function $Q\left(  x,t\right)  $ is (see
Eqs.~(\ref{qu}) and (\ref{ini}))
\begin{equation}
Q\left(  x,0\right)  =W\left(  x,0\right)  \eta\left(  0\right)  =\pm
\delta\left(  x-x_{0}\right)  . \label{triv}%
\end{equation}
Let us consider the potential profiles $U\left(x \right)  \pm
V\left( x\right)  $ with a wall at $x\rightarrow-\infty$ and a
sink at $x\rightarrow +\infty$ (see Fig.~\ref{fig1}). The
potential profile $U\left(  x\right) +V\left(  x\right)  $
corresponds to a metastable state, and $U\left( x\right)  -V\left(
x\right)  $ corresponds to an unstable one.
Thus, we investigate the system with randomly switching metastable
state.

The nonlinear relaxation time (NLRT) for the state located in the interval
$\left(  L_{1},L_{2}\right)  $ is defined as follows \cite{Agu99}
\begin{equation}
\tau\left(  x_{0}\right)  =\int_{0}^{\infty}dt\int_{L_{1}}^{L_{2}}W\left(
x,t\left\vert x_{0},0\right.  \right)  dx, \label{taudef}%
\end{equation}
where $x_{0}\in\left(  L_{1},L_{2}\right)  $. The NLRT is also interpreted as
mean lifetime of Brownian particles in the interval $\left(  L_{1}%
,L_{2}\right)  $ or average residence time, because, in accordance
with Eq.~(\ref{known}), Eq.~(\ref{taudef}) can be rewritten as
conditional time average

\[
\tau\left(  x_{0}\right)  =\left\langle \int_{0}^{\infty}\theta\left(
x\left(  t\right)  -L_{1}\right)  \theta\left(  L_{2}-x\left(  t\right)
\right)  dt\left\vert x\left(  0\right)  =x_{0}\right.  \right\rangle
\]
where $\theta\left(  x\right)  $ is the step function.

Let us rewrite the definition (\ref{taudef}) in the form
\begin{equation}
\tau\left(  x_{0}\right)  =\int_{L_{1}}^{L_{2}}Y\left(
x,x_{0},0\right)  dx, \label{new-def}
\end{equation}
where $Y\left(  x,x_{0},s\right)  $ is the Laplace transform of
conditional probability density $W\left(  x,t\left\vert
x_{0},0\right.  \right)  $. After Laplace transforming
Eqs.~(\ref{two}), with initial conditions (\ref{ini}) and
(\ref{triv}), we obtain the following closed set of ordinary
differential equations
\begin{align}
DY^{\prime\prime}+\left[  U^{\prime}\left(  x\right)
Y+V^{\prime}\left( x\right)  R\right]  ^{\prime}-sY =-\delta\left(
x-x_{0}\right) ,
\nonumber\\
DR^{\prime\prime}+\left[  U^{\prime}R+V^{\prime}Y\right]  ^{\prime}-(s+2\nu)R
=\mp\delta\left(  x-x_{0}\right)  , \label{prel}%
\end{align}
where $R\left(  x,x_{0},s\right)  $ is the Laplace transform of
auxiliary function $Q\left(  x,t\right)$, defined by
Eq.~(\ref{triv}). Using the method proposed in Ref.~\cite{Mal97},
we expand the functions $sY\left( x,x_{0},s\right)  $ and
$sR\left( x,x_{0},s\right)  $ in power series in $s$
\begin{align}
sY\left(  x,x_{0},s\right)   &  =Z_{0}\left(  x,x_{0}\right)  +sZ_{1}\left(
x,x_{0}\right)  +\dots \nonumber\\
sR\left(  x,x_{0},s\right)   &  =R_{0}\left(  x,x_{0}\right)  +sR_{1}\left(
x,x_{0}\right)  +\dots \label{expand}%
\end{align}
Since all Brownian particles move to the sink located at the point $x=+\infty$
(see Fig.~\ref{fig1}) we have zero stationary probability distribution, i.e.
\[
\lim_{t\rightarrow\infty}W\left(  x,t\left\vert x_{0},0\right.  \right)
=\lim_{s\rightarrow0}sY\left(  x,x_{0},s\right)  =0.
\]
As a consequence, in expansions (\ref{expand}) $Z_{0}\left(
x,x_{0}\right) =0$, $R_{0}\left(  x,x_{0}\right)  =0$, and the
definition (\ref{new-def}) becomes
\begin{equation}
\tau\left(  x_{0}\right)  =\int_{L_{1}}^{L_{2}}Z_{1}\left(  x,x_{0}\right)
dx. \label{tau}%
\end{equation}
Substituting the expansions (\ref{expand}) in Eqs.~(\ref{prel}) and equating
the terms without $s$, we obtain the following set of equations for the
functions $Z_{1}\left(  x,x_{0}\right)  $ and $R_{1}\left(  x,x_{0}\right)  $
\begin{align}
DZ_{1}^{\prime\prime}+\left[  U^{\prime}\left(  x\right)
Z_{1}+V^{\prime }\left(  x\right)  R_{1}\right]  ^{\prime}  &
=-\delta\left(  x-x_{0}\right) ,
\nonumber\\
DR_{1}^{\prime\prime}+\left[  U^{\prime}R_{1}+V^{\prime}Z_{1}\right]
^{\prime}-2\nu R_{1}  &  =\mp\delta\left(  x-x_{0}\right)  . \label{near}%
\end{align}

Because of the reflecting boundary at $x=-\infty$, the probability
flow equals zero at this point, and from Eqs.~(\ref{two}) we have
\begin{align}
\left[  D\frac{\partial W}{\partial x}+U^{\prime}\left(  x\right)
W+V^{\prime}\left(  x\right)  Q\right]  _{x=-\infty}  &  =0,\nonumber\\
\left[  D\frac{\partial Q}{\partial x}+U^{\prime}\left(  x\right)
Q+V^{\prime}\left(  x\right)  W\right]  _{x=-\infty}  &  =0. \label{refl}%
\end{align}
Making the Laplace transform of Eqs.~(\ref{refl}) and substituting
the expansions (\ref{expand}), we obtain the following conditions
for the functions $Z_{1}\left(  x,x_{0}\right)  $ and $R_{1}\left(
x,x_{0}\right)  $
\begin{align}
\left[  DZ_{1}^{\prime}+U^{\prime}\left(  x\right)  Z_{1}+V^{\prime}\left(
x\right)  R_{1}\right]  _{x=-\infty}  &  =0,\nonumber\\
\left[  DR_{1}^{\prime}+U^{\prime}\left(  x\right)  R_{1}+V^{\prime}\left(
x\right)  Z_{1}\right]  _{x=-\infty}  &  =0. \label{one}%
\end{align}
By integrating the system (\ref{near}) from $-\infty$ to $x$, with
boundary conditions~(\ref{one}), we obtain the following closed
set of integro-differential equations for the functions
$Z_{1}\left(x,x_{0}\right)$ and $R_{1}\left(x,x_{0}\right)$
\begin{eqnarray}
DZ_{1}^{\prime}+U^{\prime}\left(  x\right)  Z_{1}+V^{\prime}\left(
x\right)
R_{1} =-\theta\left(  x-x_{0}\right) ,\nonumber\\
DR_{1}^{\prime}+U^{\prime}R_{1}+V^{\prime}Z_{1} =2\nu\int_{-\infty}%
^{x}R_{1}dy\mp\theta\left(  x-x_{0}\right)  .
\label{finally}%
\end{eqnarray}
These general equations allow to calculate the NLRT for potential
profiles above-defined. We may consider two mean lifetimes
$\tau_{+}\left(  x_{0}\right)  $ and $\tau _{-}\left(
x_{0}\right)  $, depending on the initial configuration of the
randomly switching potential profile $\Phi(x,0)$: $U\left(
x\right) +V\left(  x\right)  $ or $U\left( x\right)  -V\left(
x\right)  $. The NLRT (\ref{tau}) is equal to $\tau_{+}\left(
x_{0}\right)  $, when we take the sign "$-$" in the second
equation of system (\ref{finally}), and vice versa for
$\tau_{-}\left(  x_{0}\right)  $.

\section{Lifetimes for piece-wise linear potential}

\begin{figure}[ptb]
\includegraphics{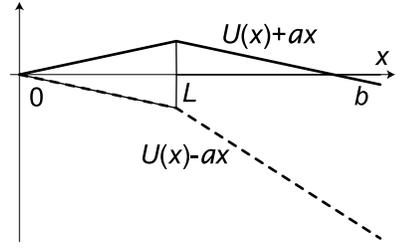}\caption{Switching piece-wise linear potential.}%
\label{fig2}%
\end{figure}

Let us consider a piece-wise linear potential profile (see
Fig.~\ref{fig2}) with $V\left(  x\right)  =ax$ ($x>0$, $0<a<k$)
and
\begin{align}
U(x)=\left\{
\begin{array}
[c]{ll}%
+\infty, & x<0\\
0, & 0\leq x\leq L\\
k\left(  L-x\right)  , & x>L
\end{array}
\right.  .\label{piecewise}
\end{align}
Hereafter we shall analyze the average residence time $\tau_{-}\left(
0\right)  $ from the interval $\left(  L_{1}=0,L_{2}=b\right)  $ with $b>L$,
which is finite in deterministic case. We consider the initial position of all
Brownian particles at the origin, i.e. $x_{0}=0$. The potential profile
$U\left(  x\right)  +ax$ corresponds to metastable state and $U\left(
x\right)  -ax$ corresponds to unstable one, as indicated in Fig.~\ref{fig2}.
After substituting the potential (\ref{piecewise}) and $V\left(  x\right)
=ax$ in Eqs.~(\ref{finally}) and choosing the sign "+" we arrive at
\begin{align}
DZ_{1}^{\prime}-k\cdot\theta\left(  x-L\right)  Z_{1}+aR_{1}  &
=-1,\nonumber\\
DR_{1}^{\prime}-k\cdot\theta\left(  x-L\right)  R_{1}+aZ_{1}  &  =1+2\nu
\int_{0}^{x}R_{1}dy. \label{area1}%
\end{align}
We solve the set of differential equations (\ref{area1}) in the
regions $0<x<L$ and $x>L$ separately, and then use the continuity
conditions at the point $x=L$
\begin{equation}
Z_{1}\left\vert _{L-0}\right.  =Z_{1}\left\vert _{L+0}\right.  ,\quad
R_{1}\left\vert _{L-0}\right.  =R_{1}\left\vert _{L+0}\right.  . \label{cont}%
\end{equation}

For $0<x<L$, the solutions of Eqs.~(\ref{area1}) read
\begin{eqnarray}
 Z_{1}\left(  x\right) &=&c_{1}\left(  \cosh{\gamma x}+\frac{2\nu D}{a^{2}%
}\right)  +c_{2}\sinh{\gamma x}+\frac{1}{a}-\frac{2\nu
x}{\gamma^{2}D^{2}},\nonumber\\
R_{1}\left(  x\right) &=& -\frac{\gamma D}{a}\left(
c_{1}\sinh{\gamma
x}+c_{2}\cosh{\gamma x}\right)  -\frac{a}{\gamma^{2}D^{2}},\label{first}%
\end{eqnarray}
where $Z_{1}\left(  x\right)  \equiv Z_{1}\left(  x,0\right)  $, $R_{1}\left(
x\right)  \equiv R_{1}\left(  x,0\right)  $ and
\begin{equation}
\gamma=\sqrt{\frac{a^{2}}{D^{2}}+\frac{2\nu}{D}}. \label{gama}%
\end{equation}
The finite solutions of Eqs.~(\ref{area1}) in the interval $(L,+\infty)$ are
\begin{align}
Z_{1}\left(  x\right)   &  =c_{3}\cdot e^{\mu\left(  x-L\right)  }+\frac{1}%
{k},\nonumber\\
R_{1}\left(  x\right)   &  =\frac{c_{3}\left(  k-\mu D\right)  }{a}\cdot
e^{\mu\left(  x-L\right)  }, \label{second}%
\end{align}
where
\begin{align}
& \mu =\frac{2k}{3D}\left[  1+\sqrt{1+3\frac{\gamma^{2}D^{2}}{k^{2}}}%
\cos\left(  \frac{\theta+2\pi}{3}\right)  \right]  ,\nonumber\\
& \cos\theta =-\frac{1+9\left(  \nu D-a^{2}\right)  /k^{2}}{\left[
1+3\gamma^{2}D^{2}/k^{2}\right]  ^{3/2}} \label{mu}%
\end{align}
is the negative root of the following cubic equation
\begin{equation}
\lambda{\left(  \lambda-\frac{k}{D}\right)  }^{2}-\gamma^{2}\lambda+\frac{2\nu
k}{D^{2}}=0. \label{char}%
\end{equation}
Substituting the solutions (\ref{first}) and (\ref{second}) in the
continuity conditions (\ref{cont}) and in the second equation
(\ref{area1}), we obtain, after rearrangements, the following
compact system of algebraic equations for unknown constants
$c_{1},c_{2},c_{3}$
\begin{align}
& c_{1}\cosh{\gamma L}+c_{2}\sinh{\gamma L}+c_{3}\left( \frac{2\nu k}
{\mu \Gamma^2}-1\right) = 0,\nonumber\\
& c_{1}\sinh{\gamma L}+c_{2}\cosh{\gamma L}+c_{3}\frac{k-\mu
D}{\Gamma} = -\frac{a^{2}}{\Gamma^3},\label{al-sys}\\
& c_{1}-c_{3}\frac{ka^{2}}{\mu \Gamma^2 D}=\frac{a h}{2\nu D}
,\nonumber
\end{align}
where

\begin{eqnarray}
\Gamma &=& \gamma D,\nonumber \\
 h &=& \frac{a}{k}+\frac{2\nu
aL}{\gamma ^2}-1.
\label{Gamma_h}
\end{eqnarray}
The solutions of Eqs. (\ref{al-sys}) are

\begin{widetext}
\begin{eqnarray}
c_{1}&=&\frac{ah}{2\nu D}+\frac{a^{3}k\left(  2\nu D a\sinh{\gamma
L} - h \Gamma ^3 \right) }{2\nu\Gamma^3 D\left[ a^{2}k+D\left(
2\nu k-\mu\Gamma^2 \right)  \cosh{\gamma L+}\mu\Gamma D\left( \mu
D-k\right)  \sinh{\gamma L}\right] },
\nonumber \\
c_{2}&=&\frac{a\left\{\left( \mu\Gamma^2 -2\nu k\right) \left(
2\nu D a+h\Gamma^3\sinh{\gamma L}\right)  +\left[  h\mu\Gamma^{4}%
\left(  k-\mu D\right)  -2\nu a^{3}k\right]  \cosh{\gamma
L}\right\} }{2\nu\Gamma^3 \left[ a^{2}k+D\left( 2\nu k-\mu\Gamma^2
\right)  \cosh{\gamma L+}\mu\Gamma D\left( \mu D-k\right)
\sinh{\gamma L}\right] },
\label{c-coefficients} \\
c_{3}&=&\frac{\mu a \left(  2\nu D a\sinh{\gamma L} - h \Gamma ^3
\right)}{2\nu\Gamma\left[  a^{2}k+D\left( 2\nu k-\mu\Gamma^2
\right)  \cosh{\gamma L+}\mu\Gamma D\left( \mu D-k\right)
\sinh{\gamma L}\right]  }. \nonumber
\end{eqnarray}
\end{widetext}
Substituting the expressions \ref{first}) and (\ref{second}) of
the function $Z_{1}\left( x\right) $ in Eq.~(\ref{tau}), and using
Eqs.~(\ref{c-coefficients}) , we obtain finally the following
result for mean lifetime

\begin{widetext}
\begin{align}
&\tau_{-}\left(  0\right)  =\frac{b}{k}+\frac{\nu
L^{2}}{\Gamma^{2}} + \nonumber \\
&\frac{a}{2\nu\Gamma^{4}}\cdot(\frac{ D \Gamma\left\{  2\nu
a\left[ \Gamma^{2}\left( e^{\mu\left(  b-L\right)  }-1\right)
+2\nu kL\right]  +h\Gamma^{2}\left( 2\nu k-\mu\Gamma^{2}\right)
\right\} \sinh{\gamma L}-2\nu D a \left(
a^{2}k+\mu\Gamma^{2}D-2\nu kD\right) }{a^{2}k+D\left(  2\nu
k-\mu\Gamma ^{2}\right) \cosh{\gamma L+}\mu\Gamma D\left(  \mu
D-k\right)  \sinh{\gamma L}}+ \nonumber
\\
&\frac{D \left[  h\mu \Gamma^{4}\left( \mu D-k\right)  +2\nu a
\left( a^{2}k+\mu\Gamma^{2}D-2\nu kD\right)  \right] \cosh{\gamma
L}-h\Gamma^{4} \left[ \Gamma^{2}\left( e^{\mu\left( b-L\right)
}-1\right) +2\nu kL +\mu D(\mu D -k ) \right]}{a^{2}k+D\left( 2\nu
k-\mu\Gamma^{2}\right) \cosh{\gamma L+}\mu\Gamma D\left(  \mu
D-k\right) \sinh{\gamma L}}).
 \label{main}
\end{align}
\end{widetext}

Equation (\ref{main}) is exact, and was derived without any
assumptions on the white noise intensity $D$ and on the mean rate
of flippings $\nu$.

\section{Condition to observe noise enhanced stability (NES)}

Because of the complicated expression (\ref{main}) of the mean
lifetime, we analyze the limiting cases of very large and very
small noise intensities. Using the approximate estimations for
small parameters $\gamma$ and $\mu$ in the limit
$D\rightarrow\infty$
\[
\gamma\simeq\sqrt{\frac{2\nu}{D}}\left(  1+\frac{a^{2}}{4\nu D}\right)
,\quad\mu\simeq-\sqrt{\frac{2\nu}{D}}\left(  1-\frac{k}{2\sqrt{2\nu D}%
}\right)  ,
\]
obtained from Eqs.~(\ref{gama}) and (\ref{mu}), we find from
Eq.~(\ref{main})
\begin{equation}
\tau_{-}\left(  0\right)  =\frac{b}{k}+\frac{L^{2}}{2D}\left[
1-\frac {bq\left(  1-q\right)  }{\omega L}\right]  +o\left(
\frac{1}{D}\right) \ .
\label{big}%
\end{equation}
Here $\omega=\nu L/k$ and $q=a/k$ are dimensionless parameters.
The parameter $q$ quantifies the degree of potential flatness
after the point $L$ (see Fig.~\ref{fig2}). Under very large noise
intensity $D$, Brownian particles "do not see" the fine structure
of potential profile and move as in the fixed potential $-kx$.
Therefore NLRT decreases with noise intensity, tending to the
value $b/k$ as follows from Eq.~(\ref{big}).

The NES phenomenon should be searched in opposite limiting case of
very slow diffusion $(D\rightarrow0)$ \cite{Agu01,Man96,Spa04}.
The approximate expressions, obtained in this limit, for
parameters $\gamma, \Gamma$ and $\mu$ are
\begin{eqnarray}
\gamma &  \simeq &\frac{a}{D}\left(  1+\frac{\nu D}{a^{2}}\right)
, \quad \Gamma \simeq a \left(  1+\frac{\nu D}{a^{2}}\right), \nonumber\\
\mu &  \simeq & -\frac{2\omega}{L\left(  1-q^{2}\right)  }\left[
1-\frac{2\omega D\left(  1+q^{2}\right)  }{kL\left( 1-q^{2}\right)
^{2}}\right] \ .\label{para}
\end{eqnarray}
Substituting Eqs.~(\ref{para}) in Eq.~(\ref{main}), and retaining
the terms up to first order in $D$, we obtain the following
expression for NLRT at small noise intensity
\begin{equation}
\tau_{-}\left(  0\right)  =\tau_{0}+\frac{D}{a^{2}}\cdot f(q,\omega
,s)+o\left(  D\right) \ . \label{simp}%
\end{equation}
Here
\begin{widetext}
\begin{equation}
f(q,\omega,s) =  \frac{3q^2+4q-5}{2(1-q^2)} + 2\omega
\frac{3q^2+q-3}{q(1-q^2)} - \frac {2\omega^2}{q^2}+ se^{-s}\frac
{q^3(1+q^2)}{(1+q)(1-q^2)}+ \left( 1-e^{-s}\right) \frac
{q(1-q^2-2q^3)}{2(1-q^2)} \label{grad}
\end{equation}
\end{widetext}
and
\begin{equation}
\tau_{0}=\frac{2L}{a}+\frac{\nu L^{2}}{a^{2}}+\frac{b-L}{k}-\frac{q(1-q)}%
{2\nu}\left(  1-e^{-s}\right)  \label{tau0}%
\end{equation}
is the mean lifetime in the absence of white Gaussian noise $(D=0)$. In
Eqs.~(\ref{grad}) and (\ref{tau0}) new dimensionless parameter
\[
s=\frac{2\omega\left(  b/L-1\right)  }{1-q^{2}}
\]
is introduced.

For very slow switchings $\nu\rightarrow0$, we find from
Eq.~(\ref{tau0})
\begin{equation}
\tau_{0}\left(  \nu\rightarrow0\right)  =\frac{2L}{a}+\frac{b-L}{k+a},
\label{tau+0}%
\end{equation}
which is different from the deterministic time
\begin{equation}
\tau_{d}=\tau_{0}\left(  \nu=0\right)  =\frac{L}{a}+\frac{b-L}{k+a}.
\label{tau-d}%
\end{equation}
Difference between the results (\ref{tau+0}) and (\ref{tau-d}) is due to a
nonzero probability of one switching within the deterministic time interval
$\left(  0,\tau_{d}\right)  $ in the case $\nu\rightarrow0$.

\begin{figure}[ptb]
\includegraphics{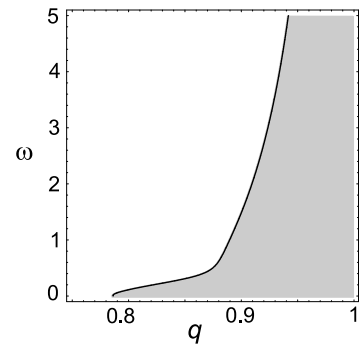}\caption{Shaded area is the parameter region on the
plane $(q,\omega)$ where NES effect can be observed. Here
$\omega=(\nu L)/k$, $q=a/k$, and $b=2L$.}%
\label{fig3}%
\end{figure}

The condition to observe the NES effect can be expressed by the inequality
\begin{equation}
f\left(  q,\omega,s\right)  >0. \label{NES}%
\end{equation}
Let us analyze the structure of NES region on the plane $\left(
q,\omega\right)  $ from Eqs.~(\ref{grad}) and (\ref{NES}). At very slow and
fast flippings we obtain
\begin{eqnarray}
q >\frac{\sqrt{19}-2}{3}\simeq0,7863,& &\quad\omega\rightarrow0\nonumber\\
\omega <\frac{q(3q^{2}+q-3)}{1-q^{2}},& &
\quad\omega\rightarrow\infty.
\label{two-in}%
\end{eqnarray}
In Fig.~\ref{fig3} we show the NES region (shaded area) on the
plane $\left( q,\omega\right)  $ for $b/L=2$, obtained from
inequality (\ref{NES}).
The NES effect occurs at $q\simeq1$, i.e. at very small steepness
$k-a=k\left(  1-q\right)  $ of the reverse potential barrier for
the metastable state. For this potential profile, a small noise
intensity can return particles into potential well, after they
crossed the point $L$. Then Brownian particles stay for long time
in the metastable state. This means that, for a fixed mean
flipping rate, the NES effect increases when $q \rightarrow 1$.
For fixed parameter $q$ the effect increases when
$\omega\rightarrow0$, because Brownian particles have enough time
to move back into potential well.

\begin{figure}[ptb]
\includegraphics{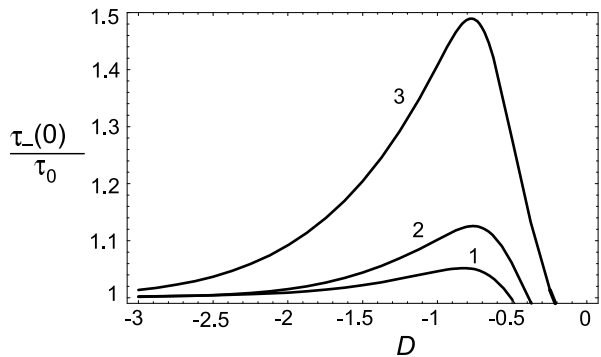}\caption{Semilogarithmic plot
of the normalized mean lifetime $\tau_{-}\left(  0\right)
/\tau_{0}$ vs the white noise intensity $D$ for three values of
the dimensionless mean flipping rate $\omega=\nu L/k$: $0.1$
(curve $1$), $0.05$ (curve $2$), $0.01$ (curve $3$). Parameters
are $L=1$, $k=1$, $b=2$, and $a=0.995$.}
\label{fig4}
\end{figure}

In Fig.~\ref{fig4} we show the plots of the normalized mean
lifetime $\tau _{-}\left(0\right) /\tau_{0}$, Eq. (\ref{main}), as
a function of the noise intensity $D$ for three values of the
dimensionless mean flipping rate $\omega=\nu L/k$: $0.01$, $0.05$,
$0.1$.
The maximum value of the NLRT and the range of noise intensity
values, where NES effect occurs, increases when $\omega$
decreases.

\begin{figure}[ptb]
\includegraphics{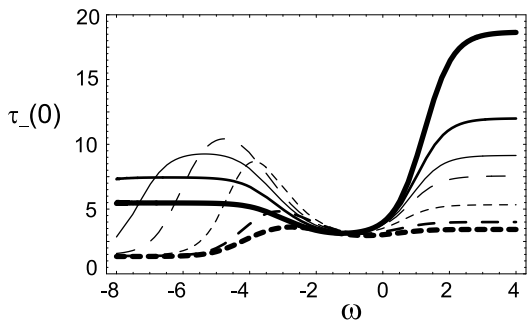}\caption{Semilogarithmic plot of the mean lifetime
$\tau_{-}\left(  0\right)  $ vs the dimensionless mean flipping
rate $\omega=\nu L/k$ for seven noise intensity values.
Specifically from top to bottom on the right side of the figure:
$D = 0.03,0.05,0.07,0.09,0.15,0.25,0.35$. The other parameters are
the same as in Fig.~\ref{fig4}.} \label{fig5}
\end{figure}

By using exact Eq.~(\ref{main}) we have also investigated the
behavior of the mean lifetime $\tau_{-}\left( 0\right)  $ as a
function of switchings mean rate $\nu$. In Fig.~\ref{fig5} we plot
this behavior for seven values of noise intensity.
At very slow flippings $(\nu\rightarrow0)$ we obtain
\begin{equation}
\tau_{-}\left(  0\right)  \simeq\tau_{d}-\frac{D\left(  1-e^{-aL/D} \right)
}{a^{2}\left(  1+q\right)  }, \label{slow}%
\end{equation}
i.e. the NLRT of the fixed unstable potential $U\left(  x\right)  -ax$. While
for very fast switchings $(\nu\rightarrow\infty)$ we obtain
\begin{equation}
\tau_{-}\left(  0\right)  \simeq\frac{b}{k} + \frac{L^{2}}{2D} , \label{fast}%
\end{equation}
i.e. the mean lifetime for average potential $U\left(  x\right) $.
All limiting values of the NLRT expressed by Eqs.~(\ref{slow}) and
(\ref{fast}) are shown in Fig.~\ref{fig5}. At intermediate rates
the escape from the metastable state exhibits a minimum at
$\omega=0.1$, which is the signature of resonant activation (RA)
phenomenon \cite{Doe92,Man00,Bie93,Iwa96}.

Moreover, in Fig. \ref{fig5} we observe a new resonant-like
behavior for the NLRT as a function of mean fluctuation rate of
potential. The NLRT exhibits a \emph{maximum} between the slow
limit of potential fluctuations (static limit) and the RA minimum.
This maximum occurs for a value of the barrier fluctuation rate on
the order of the inverse of the time $\tau_{up}\left(  D\right)  $
required to escape from the metastable fixed configuration
\begin{equation}
\tau_{up}\left(  D\right)  = \frac{b-L}{k-a} - \frac{L}{a} + \frac{D\left(
e^{aL/D}-1\right)  }{a^{2}\left(  1-q\right)  }. \label{upper-tau}%
\end{equation}
This suggests that, the enhancement of stability of metastable
state is strongly correlated with the potential fluctuations, when
the Brownian particle "sees" the barrier of the metastable state
\cite{Agu01,Man96,Spa04}.

\section{Conclusions}

We have investigated the nonlinear relaxation time for
one-dimensional system with additive white Gaussian noise, and
potential profile switching between two configurations, due to a
Markovian dichotomous noise. From the general equations
(\ref{finally}), we provide exact expression of the mean lifetime
for piece-wise linear potential, for arbitrary noise intensity,
and arbitrary fluctuation rate of the potential. We find the noise
enhanced stability and the resonant activation phenomena in the
system investigated. We obtained analytically the region on the
$(q,\omega)$ plane, where the NES effect can be observed.
Moreover, when we fix white noise intensity $D$, flatness $q$, and
vary switchings mean rate $\nu$, we can observe new resonant-like
behavior of the mean lifetime, which is related to the NES
phenomenon. The NLRT shows a maximum as a function of the mean
flipping rate of potential, with the NES effect strongly
correlated with the potential fluctuations. The general equations
derived in this paper enable us to perform the analysis of the NES
effect conditions in physical systems with more complex potential
profiles.

\section*{ACKNOWLEDGMENTS} This work has been supported by INTAS
Grant 2001-0450, MIUR, INFM, by Russian Foundation for Basic
Research (project 02-02-17517), by Federal Program "Scientific
Schools of Russia" (project 1729.2003.2), and by Scientific
Program "Universities of Russia" (project 01.01.020).

\appendix

\section*{Appendix}

Upon differentiation of Eq.~(\ref{known}) on $t$, we obtain
\begin{equation}
\frac{\partial W}{\partial t}=-\frac{\partial}{\partial x}\left\langle \dot
{x}\left(  t\right)  \delta\left(  x-x\left(  t\right)  \right)  \right\rangle
. \label{W-1}%
\end{equation}
Substituting $\dot{x}\left(  t\right)  $ from Eq.~(\ref{Lang}),
and using the definition (\ref{qu}) of auxiliary function, we can
rewrite Eq.~(\ref{W-1}) as
\begin{align}
\frac{\partial W}{\partial t}  &  =\frac{\partial}{\partial x}\left[
U^{\prime}(x)W\right]  +\frac{\partial}{\partial x}\left[  V^{\prime
}(x)Q\right] \nonumber\\
&  -\frac{\partial}{\partial x}\left\langle \xi\left(  t\right)  \delta\left(
x-x\left(  t\right)  \right)  \right\rangle . \label{W-2}%
\end{align}
To obtain the evolution of function $Q\left(x,t\right)$, we use
the Shapiro-Loginov's formula for Markovian dichotomous noise
\cite{Sha78}
\begin{equation}
\frac{d}{dt}\left\langle \eta\left(  t\right)  R_{t}\left[  \eta\right]
\right\rangle =-2\nu\left\langle \eta\left(  t\right)  R_{t}\left[
\eta\right]  \right\rangle +\left\langle \eta\left(  t\right)  \dot{R}%
_{t}\left[  \eta\right]  \right\rangle , \label{Sha-Log}%
\end{equation}
where $R_{t}\left[\eta\right]$ is an arbitrary functional
depending on the history of random process $\eta\left( \tau\right)
$, $0 \leq \tau\leq t$. Replacing $R_{t}\left[  \eta\right]$ with
$\delta\left(x-x\left( t\right)\right)$ in Eq.~(\ref{Sha-Log}),
using Eq.~(\ref{Lang}) and taking into account that
$\eta^{2}\left(  t\right) = 1$, we arrive at
\begin{align}
\frac{\partial Q}{\partial t}  &  =-2\nu Q+\frac{\partial}{\partial x}\left[
U^{\prime}(x)Q\right]  +\frac{\partial}{\partial x}\left[  V^{\prime
}(x)W\right] \nonumber\\
&  -\frac{\partial}{\partial x}\left\langle \xi\left(  t\right)  \eta\left(
t\right)  \delta\left(  x-x\left(  t\right)  \right)  \right\rangle .
\label{Q-2}%
\end{align}

To split the functional averages in Eqs.~(\ref{W-2}) and
(\ref{Q-2}), we use the Furutsu-Novikov's formula for white
Gaussian noise $\xi\left( t\right)  $ \cite{Nov65}
\begin{align}
\left\langle \xi\left(  t\right)  F_{t}\left[  \xi\right]  \right\rangle  &
=\int_{0}^{t}\left\langle \xi\left(  t\right)  \xi\left(  \tau\right)
\right\rangle \left\langle \frac{\delta F_{t}\left[  \xi\right]  }{\delta
\xi\left(  \tau\right)  }\right\rangle d\tau\nonumber\\
&  =D\left\langle \frac{\delta F_{t}\left[  \xi\right]  }{\delta\xi\left(
t\right)  }\right\rangle , \label{F-N}%
\end{align}
where $F_{t}\left[\xi\right]$ is an arbitrary functional of
$\xi\left( t\right)$. Replacing sequentially $F_{t}\left[
\xi\right]$ with $\delta\left(  x-x\left(t\right) \right)$ and
with $\eta\left(t\right) \delta\left(  x-x\left(  t\right)
\right)$ in Eq.~(\ref{F-N}), and taking into account that, in
accordance with Eq.~(\ref{Lang}), $\delta x\left( t\right)
/\delta\xi\left(t\right) =1$, we find
\begin{align}
\left\langle \xi\left(  t\right)  \delta\left(  x-x\left(  t\right)  \right)
\right\rangle  &  =-D\frac{\partial W}{\partial x},\nonumber\\
\left\langle \xi\left(  t\right)  \eta\left(  t\right)  \delta\left(
x-x\left(  t\right)  \right)  \right\rangle  &  =-D\frac{\partial Q}{\partial
x}. \label{ave}%
\end{align}
Substituting the expressions (\ref{ave}) in Eqs.~(\ref{W-2}) and
(\ref{Q-2}), we obtain the desired closed set of equations
(\ref{two}).

\end{document}